\documentclass[referee]{raa}            

\usepackage{graphicx,times}             
\usepackage{natbib}
\usepackage{amssymb,amsmath}
\bibpunct{(}{)}{;}{a}{}{,}

\usepackage{tabularx}
\usepackage{multirow}
\usepackage{ulem}
\usepackage{cancel} 
\usepackage[utf8]{inputenc}

\usepackage[a4paper]{geometry}
\usepackage[pagebackref=true]{hyperref}

\begin{document}
	
	\title{A Challenge of Developing a Classifier for Multi-Band Classification of Variable Stars}
	
	\volnopage{Vol.0 (20xx) No.0, 000--000}      
	\setcounter{page}{1}          
	
	\author{Xiao-Hui Xu  
		\inst{1,2}
		\and Qing-Feng Zhu
		\inst{1,2,3}
		\and Xu-Zhi Li
		\inst{1,2, 4, 5}
		\and Hang Zheng
		\inst{1,2}
		\and Jin-Sheng Qiu
		\inst{1,2}
	}
	
	\institute{CAS Key Laboratory for Research in Galaxies and Cosmology, Department of Astronomy, University of Science and Technology of China, Hefei 230026, China; {\it xiaohuixu@mail.ustc.edu.cn zhuqf@ustc.edu.cn}\\
		\and
		School of Astronomy and Space Sciences, University of Science and Technology of China, Hefei 230026, China\\
		\and
		Deep Space Exploration Laboratory, Hefei 230088, China\\
        \and
        School of Mathematics and Physics, Anqing Normal University, Anqing 246133, China\\
        \and
        Institute of Astronomy and Astrophysics, Anqing Normal University, Anqing 246133, China\\
		\vs\no
		{\small Received 20xx month day; accepted 20xx month day}}
	
	\abstract{Variable stars play a very important role in our understanding of the Milky Way and the universe. In recent years, many survey projects have generated a large amount of photometric data, necessitating classifiers that can quickly identify various types of variable stars. However, obtaining these classifiers often requires substantial manpower and computational resources. To conserve these resources, it would be best to have a classifier that can be used across surveys. We explore the possibility that a classifier created in one optical band can also work in other bands, likely from different survey facilities. We construct a random forest classifier based on photometric data in ASAS-SN V-band and OGLE I-band, and apply the classifier on ASAS-SN V-band and ZTF r-band light curves of variable star samples. We explore the classification differences of using the magnitude light-curves or the normalized flux light-curves, the periods derived from single band light-curves or the periods derived from the multi-band combined light-curves, and with or without color-related features. We find it feasible to develop a classifier capable of working in both V and r bands for certain types of variable stars, such as RRAB variables. For other types of variable stars, like Cepheids, the classifier is unable to make accurate identifications. }
 
        \authorrunning{X.-H. Xu, Q.-F. Zhu, X.-Z. Li, H. Zheng, J.-S. Qiu} 
        \titlerunning{The Challenge of Developing a Classifier for Multi-Band Classification of Variable Stars}  
	
        \maketitle
	
	%
	%
	\section{Introduction}           
	\label{sec:packages1}
	
	Variable stars are a large and important branch of time-varying sources that help us understand the Milky Way and the universe. Classical Cepheids (DCEPs) are the extra-galactic distance ladder and eventually lead to the determination of the Hubble constant (H$_{0}$) thanks to the well-defined period-luminosity and period-Wesenheit relations \citep{Riess2011, Riess2019}. The classic Cepheids discovered by \cite{Chen2019} created the first intuitive 3D map of the Milky Way's warping precession. RR Lyrae stars are powerful probes of the disk and halo of the Galaxy and are useful to explore the chemical and dynamical evolution of the Galaxy \citep{Medina2022, Hansen2011, Hansen2016, Li&Binney2022}. By tracing the radial velocity of eclipsing binary stars, physical parameters (such as the radius and mass) of the stars in the system can be derived \citep{Torres2010}. \cite{PamosOrtega2023} dated young open clusters (Trumpler 10 and Praesepe) using asteroseismology studies of $\delta$ Scuti stars. 

	Over the past decades, numerous variable star classifiers have been developed to facilitate the rapid identification and classification of variable stars. \cite{Kim2014} selected 28,392 light curves from the EROS-2 LMC database and extracted 22 variability features of light curves to train a random forest classifier. They used EROS B$_{E}$-band light curves (either raw- or phase-folded) to derive these features, except for \emph{J}, \emph{Q$_{3-1\vert{B-R}}$} and \emph{$\eta$$_{B-R}$}, which needs both B$_{E}$-and R$_{E}$-band light curves. \cite{KimBailer-Jones2016} provided a Python package “UPSILoN” containing a pre-trained random forest classifier trained on the light curves of 143,923 periodic variables from the OGLE I-band and EROS-2 B$_{E}$-band \citep{Tisserand2007} databases, They choosed 16 features that were relatively survey-independent, i.e. excluded the ones based colors. \cite{Xu2022} selected 241,154 periodic variable stars from ASAS-SN \citep{Shappee2014,Jayasinghe2018a,Jayasinghe2018b} V-band and OGLE \citep{Udalski2003} I-band. They extracted 17 features to train the random forest classifier. \cite{Szklenar2022} created a Multiple-Input Neural Network, consisting of Convolutional and Multi-layer Neural Networks. The network was trained and tested on OGLE-III I-band phase-folded light curves, while taking also into account additional numerical information (e.g. period, reddening-free brightness) to differentiate visually similar light curves. \cite{Abdollahi2023} used the Hierarchical Classification technique and trained the Neural Networks based on the I-band photometry of variables in OGLE-IV database. They folded the light curves by using the specific periods provided by the OGLE catalogue and tried to bin and normalize them. They used 50 values of the binned magnitudes and the period of each light curve as the initial data. All these classifiers are trained by using features extracted from single-band or multi-band light curves, with color-related information sometimes used as additional input. 
 
    However, building effective variable star classifiers for every band and survey, especially when dealing with large-scale data and complex deep learning models, is a highly resource-intensive process. It involves multiple steps such as visual classification, manual source screening \citep{Jayasinghe2018b}, data cleaning, feature construction, model training, hyperparameter tuning, validation evaluation, and iterative optimization \citep{2024A&C....4800850C,2024arXiv241022869E}, each requiring substantial computing resources and time.
 
    Certain types of variable stars exhibit a degree of consistency in different bands. Cepheid variables show stable Period-Luminosity (PL) relations and Period-Wesenheit (PW) relations in multiple bands. These relations are particularly evident in the optical bands but can also be extended to the near-infrared bands \citep{Ngeow2022a,Ngeow2022b,Munoz2024}. Similarly, RR Lyrae stars demonstrate a consistent relation between their periods and metallicity across different bands \citep{Bhardwaj2024,Narloch2024}. $\delta$ Scuti stars display similar frequency modes across different bands. \cite{Jia2024} found 2254 multi-mode $\delta$ Scuti stars from the Zwicky Transient Facility Survey. And 1254 objects showed reliable periods in both bands. 
    These findings lead us to consider developing a variable star classifier that is both efficient and resource-saving across multiple bands. To test this idea, we should build a classifier based on photometric data from one band and then compare the classification consistency on photometric data from the second and third bands. However, we do not have suitable data from three bands to perform this work, we instead use photometric data from ASAS-SN V band and ZTF DR2 r band. Since the light curves from the two bands satisfy the requirements for training a robust classifier and can be used to compare the classification consistency, we construct a random forest classifier primarily based on V-band light curves and apply it to classify V-band and r-band light curves. We think the classification differences between V band and r band can already tell us the limitation of the cross-band application of the classifier. We systematically compare the classification results under various preprocessing methods, including the use of magnitude light curves and normalized flux light curves, periods extracted from single-band and multi-band light curves, and the consideration of color features, thereby verifying to some extent whether different bands have varying impacts on variable star classification. 

    We introduce the variable star samples for training the random forest classifier and verifying the multi-band classification differences in Section~\ref{sec:packages2}. We show the classification results of the classifier on variable star samples in different cases in Section~\ref{sec:packages3}. We summarize and compare these results in Section~\ref{sec:packages4}and draw preliminary conclusions in Section~\ref{sec:packages5}.


	\section{DATA and METHODS}
	\label{sec:packages2}

    We select variable star samples following the method described in \cite{Xu2022}. The 220,907 V-band periodic variables with \emph{classification probability} $\geq$ 0.6 and \emph{lksl} $\leq$ 0.5 are selected from the ASAS-SN Variable Star Database \citep[The ASAS-SN Catalog of Variable Stars I - IX: ][]{Jayasinghe2018b,Jayasinghe2019b,Jayasinghe2019a,Jayasinghe2020a,Jayasinghe2021}. The \emph{lksl} (Laflfler-Kinmann string length statistic) is a statistic first used by \cite{LaflerKinman1965} to indicate the data quality of light curves. Values less than 0.5 for the \emph{lksl} suggest that  the variables have less dispersion in phased light curves, higher classification probability, and more accuracy periods. 
	

	Because samples of Cepheids, RRDs and ELL variables  meeting requirements from ASAS-SN are relatively small, we select 20,247 more periodic variables with \emph{lksl} $\leq$ 0.5 from the OGLE I-band database \citep{Soszynski2016b,Soszynski2016a,Soszynski2017a,Soszynski2017b,Soszynski2018,Soszynski2019,Soszynski2020,Pawlak2016,Udalski2018} as complement and make the size of the sample to 241,154. Although the samples from ASAS-SN and OGLE are in different bands, \cite{Xu2022} have tested that the combined samples can be used to train a classifier and obtain results with high consistence. We adopt the name scheme of ASAS-SN. Due to different name schemes between ASAS-SN and OGLE, we adopt the matching scheme in Table~\ref{tab:table1} to convert variability types. The numbers of samples of each class are shown in Table~\ref{tab:table2}.
	
	\begin{table}
		\centering
		\caption{Variability types matching scheme between OGLE and ASAS-SN catalogs.}
		\label{tab:table1}
			\begin{tabular}{lccr} 
				\hline
				Superclass & OGLE subclass & ASAS-SN subclass\\
				\hline
				Type I Cepheids\\
				& F &  \\
				& F1 & DCEP\\
				& F12 & \\
				\cline{2-3}
				& 1 &  \\
				& 12 & \\
				& 13 & \\
				& 123 &  DCEPS\\
				& 2 & \\
				& 23 & \\
				\hline
				Type II Cepheids\\
				& BLHer & \multirow{4}{*}{CWB}\\
				& BLHer\_1O & \\
				& pWVir (period $<$ 8 d) &  \\
				& WVir (period $<$ 8 d) &  \\
				\cline{2-3}
				& pWVir (period $>$ 8 d) & \multirow{2}{*}{CWA} \\
				& WVir (period $>$ 8 d) &  \\
				\cline{2-3}
				& RVTau & RVA\\
				\hline
				Eclipsing binaries\\
				& ELL & ELL\\
				\hline
				RR Lyrae\\
				& RRd & RRD\\
				\hline
			\end{tabular}
	\end{table}
 
	\begin{table}
		\centering
		\caption{The number of sources in each class from ASAS-SN and OGLE databases.}
		\label{tab:table2}
		\begin{tabular}{lcccr} 
			\hline
			Superclass & Subclass & ASAS-SN & OGLE & Total\\
			\hline
			Cepheids\\
			& CWA & 446 & 430 & 876\\
			& CWB & 371 & 463 & 834\\
			& DCEP & 1306 & 4131 & 5437\\
			& DCEPS & 430 & 2715& 3145\\
			& RVA & 58 & 261 & 319\\
			Delta Scuti\\
			& DSCT & 1916 & 0 & 1916\\
			& HADS & 3144 & 0 & 3144\\
			Eclipsing binaries\\
			& EA & 37456 & 0 & 37456\\
			& EB & 20242 & 0 & 20242\\
			& ELL & 25 & 11896 & 11921\\
			& EW & 58898 & 0 & 58898\\
			Mira\\
			& M & 9162 & 0 & 9162\\
			Rotational variables\\
			& ROT & 10049 & 0 & 10049\\
			RR Lyrae\\
			& RRAB & 25671 & 0 & 25671\\
			& RRC & 8307 & 0 & 8307\\
			& RRD & 254 & 351 & 605\\
			Semiregular variables\\
			& SR & 43172 & 0 & 43172\\
			\hline
			Total &  & 220907 & 20247 & 241154\\
			\hline
		\end{tabular}
	\end{table}

	11 features (\emph{logP}, \emph{R$_{21}$}, \emph{R$_{31}$}, \emph{R$_{41}$}, $\sigma$, \emph{ALH}, \emph{M$_{s}$}, \emph{M$_{k}$}, \emph{MAD}, \emph{IQR}, \emph{A}) were extracted for each light curve and 6 color-related features (\emph{G$_{BP}$} - \emph{G$_{RP}$}, \emph{J} - \emph{K$_{s}$}, \emph{J} - \emph{H}, \emph{W1} $-$ \emph{W2}, \emph{W$_{RP}$}, and \emph{W$_{JK}$}) were cross-matched from the Gaia Early DR3 \citep{Bailer-Jones2021}, 2MASS \citep{Skrutskie2006}, and ALLWISE \citep{Cutri2021, Wright2010} catalogs, as described in \cite{Xu2022}. The descriptions and definitions of these parameters are shown in Table~\ref{tab:table3}. 
	
	\begin{table*}
		\centering
		\caption{The 17 variability features used to train the random forest classifier and their importance.}
		\resizebox{\textwidth}{40mm}{
			\begin{tabular}{lccc} 
				\hline
				Feature & Description & Importance (\%) & Reference\\
				\hline
				\emph{logP} & Base 10 logarithm of the period & 24.2 &	-\\
				\emph{R$_{21}$} & Ratio between the amplitudes of the 2nd and 1st harmonics & 10.4	& -\\
				&   of the 6th-order Fourier model &  & \\
				\emph{$\sigma$} & Standard deviation of magnitude distribution & 8.1 & -\\
				\emph{A} &	Amplitude of the light curve (between the 5th and 95th percentiles)	& 7.0 &	-\\
				\emph{W$_{RP}$} & Absolute Wesenheit magnitude in Gaia Early DR3 \emph{G$_{RP}$}-band & 6.1 & \cite{Bailer-Jones2021}, \cite{Skrutskie2006},\\
				&  &  & \cite{Madore1982}\\
				\emph{M$_{s}$} & Skewness of the magnitude distribution	& 5.5 & -\\
				\emph{IQR}	& Difference between the 75th and 25th percentiles in magnitude	& 5.4 &	-\\
				\emph{M$_{k}$} & Kurtosis of the magnitude distribution	& 4.8 & -\\	
				\emph{W$_{JK}$} & Absolute Wesenheit magnitude in 2MASS \emph{K$_{s}$}-band & 4.4 & \cite{Bailer-Jones2021}, \cite{Skrutskie2006},\\
				&  &  &  \cite{Madore1982}\\
				\emph{MAD} & Median absolute deviation of magnitude distribution & 3.8 & -\\
				\emph{R$_{41}$} & Ratio between the amplitudes of the 4th and 1st harmonics & 3.7 & -\\
				&   of the 6th-order Fourier model &  & \\
				\emph{G$_{BP}$} $-$ \emph{G$_{RP}$} & Gaia Early DR3 \emph{G$_{BP}$} $-$ \emph{G$_{RP}$} color & 3.1	& \cite{GaiaCollaboration2021}\\
				\emph{J} $-$ \emph{H} & 2MASS \emph{J} $-$ \emph{H} color & 3.0 & \cite{Skrutskie2006}\\
				\emph{R$_{31}$} & Ratio between the amplitudes of the 3rd and 1st harmonics & 2.9 & -\\
				&   of the 6th-order Fourier model &  & \\
				\emph{J} $-$ \emph{K$_{s}$} & 2MASS \emph{J} $-$ \emph{K$_{s}$} color & 2.7 & \cite{Skrutskie2006}\\
				\emph{W1} $-$ \emph{W2} & WISE \emph{W1} $-$ \emph{W2} color & 2.3 & \cite{Cutri2013}, \cite{Wright2010}\\
				\emph{ALH}	& Ratio of magnitudes fainter or brighter than the average  & 2.3 & \cite{KimBailer-Jones2016}\\
				\hline
			\end{tabular}
		}
		\label{tab:table3}
	\end{table*}

	When using the existing random forest classifier trained by \cite{Xu2022} to investigate the multi-band variable star classification differences (seeing Appendix~\ref{sec:packages6}), we employ the period determination method as described in \cite{Xu2022}. However, in this work, we focus more on the differences of classification in different bands. Therefore, we directly use the Generalized Lomb–Scargle \citep[GLS;][]{Scargle1982,ZechmeisterKurster2009} to select the best periods without any additional processing.

    \cite{Chen2020} released the catalog and photometric data of 781,602 variable stars in ZTF DR2 g- and r-band. \cite{Xu2022} matched the coordinates with a matching radius of 5 arcsec between the ASAS-SN and ZTF DR2 catalogs, and a total of 77,262 sources were matched in V- and r-band. In this work, we use the new random forest classifiers to classify 77,262 variables in two bands. 
 
	In order to evaluate the classification differences, the classification results in the V- and r-band are regarded as set V and set r, respectively. In previous works of classification of variables, \emph{precision} represents the proportion of the true samples among classification samples; \emph{recall} represents the proportion of the samples classified correctly among true samples. \emph{F$_{1}$} is the harmonic mean of both, which evaluate the overall performance of the classifier. 
    In the current work, we define new precision and F$_{1}$ referring the definitions of \emph{precision} and \emph{F$_{1}$}. The \emph{precision1*} (i.e. $\frac{V\,\cap\,r}{V}$) represents the ratio of the number of overlapping classifications in the V and r bands to the number in V band for a particular variable type. Similarly, \emph{precision2*} (i.e. $\frac{V\,\cap\,r}{r}$) is the ratio calculated in the r band. The precision1* and precision2* show the performance of the classifiers in different bands. Ideally, two precision* should be 100\% for all variability types if the classifiers are perfectly feasible for different bands. Values less than 1 suggest certain degrees of incompetence in other bands. \emph{F$_{1}$*} (i.e. $\frac{2*\,(V\,\cap\,r)}{V+r}$) is the harmonic mean of both and indicates the classification effect of new classifiers in two bands. When the F$_{1}$* is $<$ 0.650, $\gtrsim$ 0.650 $\sim$ 0.900, $\gtrsim$ 0.900, it suggests that there are low, medium, high classification consistency or significant, moderate, minor classification differences for a specific type of variables, respectively. These criteria are totally subjective, and are not verified statistically. 

	We also define a parameter \texttt{CICD} (Comprehensive Indicator of Classification Differences) to comprehensively evaluate the classification differences of all types of variable stars in the two bands, namely:
	
	\begin{equation}
		CICD = \sum_{i=1}^{n}\frac{(V_i+r_i)*F^*_i}{2N}
	\end{equation}
	
	\noindent
	where the \emph{n} represents the numbers of variable star types, \emph{N} represents the numbers of common sources, \emph{i} represents the specific variable star type, \emph{V$_{i}$} represents the classification numbers of i-type in the V-band, \emph{r$_{i}$} represents the classification numbers of i-type in the r-band, and \emph{F$_{i}$*} represents the \emph{F$_{1}$*} of i-type. The classification numbers in each band and the classification consistency for each type of variables are taken into account. This is a weighted \emph{F$_{1}$*} for all types of variables in the samples. The lower the \texttt{CICD} is, the higher the classification differences are.

	\section{CLASSIFICATION RESULTS}
	\label{sec:packages3}
	
	\subsection{The Classification Differences Based on Normalized Flux Light-Curves}
	\label{sec:packages3.1}
	
	The variability amplitudes are used as one of features to train the random forest classifier. However, according to the results in Appendix~\ref{sec:packages6}, different magnitude variation ranges in V and r bands could cause different classification results. To avoid amplitude differences, we use the normalized flux light curves. At first, we remove the largest and smallest 1\% magnitude measurements of each light curve for 241,154 periodic variable stars. Next, we convert the magnitude light-curves into flux light-curves and each flux light-curve is divided by the maximum flux to obtain the normalized flux light-curve. The maximum flux is the average value of the largest 15\% flux values in each flux light-curve. Then we extract the features from the normalized flux light-curves and retrain the random forest classifier.
	
	When the magnitudes of variable stars are converted into normalized fluxes, the periods of a few objects can not be determined by the GLS method for unknown reasons, and a total of 235,491 objects are identified to be periodic variable stars. We extract 11 features (\emph{logP}, \emph{R$_{21}$}, \emph{R$_{31}$}, \emph{R$_{41}$}, $\sigma$, \emph{ALH}, \emph{M$_{s}$}, \emph{M$_{k}$}, \emph{MAD}, \emph{IQR}, \emph{A}) for each normalized flux light curve. For the remaining 6 color-related features (\emph{G$_{BP}$} - \emph{G$_{RP}$}, \emph{J} - \emph{K$_{s}$}, \emph{J} - \emph{H}, \emph{W1} $-$ \emph{W2}, \emph{W$_{RP}$}, and \emph{W$_{JK}$}), we keep unchanged. We dedicate 80\% of the samples to training and reserve 20\% for testing. Finally, we reconstruct the random forest classifier by using all the features. Table~\ref{tab:table4} and Table~\ref{tab:table5} show the precision, recall, and F$_{1}$ of this classifier in superclasses and subclasses. For $\delta$ Scuti, DSCT, EW, HADS, RRC, RRD, and RR Lyraes, the values of F$_{1}$ decrease by 17\%, 18\%, 9\%, 15\%, 31\%, 15\%, and 10\%, respectively. The values of F$_{1}$ increase by 7\%, 4\%, and 8\% for CWA, CWB, and RVA variables, respectively. For other types of variable stars, the values of F$_{1}$ remain largely unchanged compared to the classification performance based on magnitude light-curves \citep{Xu2022}.

	\begin{table}
		\centering
		\caption{The classification quality of superclasses based on normalized flux light-curves.}
			\begin{tabular}{lccc} 
				\hline
				Superclass & precision & recall & F$_{1}$\\
				\hline
				Cepheids & 0.888 & 0.970 & 0.927\\
				$\delta$ Scuti & 0.707 & 0.983 & 0.822\\
				Eclipsing Binaries & 0.988 & 0.923 & 0.954\\
				Mira & 0.898 & 0.993 & 0.943\\
				Rotational variables & 0.895 & 0.945 & 0.919\\
				RR Lyrae & 0.834 & 0.967 & 0.895\\
				Semiregular variables & 0.995 & 0.953 & 0.974\\
				\hline
			\end{tabular}
		\label{tab:table4}
	\end{table}


	\begin{table}
		\caption{The classification quality of subclasses based on normalized flux light-curves.}
			\centering
			\begin{tabular}{ccccc} 
				\hline
				Superclass & Subclass & precision & recall & F$_{1}$\\
				\hline
				\multirow{4}{*}{Cepheids} & CWA & 0.600 & 0.902 & 0.721\\
				& CWB & 0.525 & 0.854 & 0.651\\
				& DCEP &  0.946 & 0.857 &  0.899\\
				& DCEPS &  0.843 & 0.938 & 0.888\\
				& RVA & 0.569 &  0.984 & 0.721\\
				\cline{2-5}
				\multirow{2}{*}{$\delta$ Scuti} & DSCT & 0.639 & 0.981 & 0.774\\
				& HADS & 0.733 & 0.957 & 0.830\\
				\cline{2-5}
				\multirow{4}{*}{Eclipsing Binaries} & EA & 0.960 & 0.923 & 0.941\\
				& EB &  0.739 & 0.798 & 0.767\\
				& ELL & 0.947 & 0.969 & 0.958\\
				& EW & 0.945 & 0.804 & 0.868\\
				\cline{2-5}
				\multirow{1}{*}{Mira} & M &  0.898 & 0.993 & 0.943\\
				\cline{2-5}
				\multirow{1}{*}{Rotational variables} & ROT & 0.895 & 0.945 & 0.919\\
				\cline{2-5}
				\multirow{3}{*}{RR Lyrae} & RRAB & 0.987 & 0.971 & 0.979\\
				&RRC & 0.535 & 0.889 & 0.668\\
				& RRD &0.335 &  0.621 & 0.435\\
				\cline{2-5}
				\multirow{1}{*}{Semiregular variables} & SR & 0.995 & 0.953 & 0.974\\
				\hline
			\end{tabular}
		\label{tab:table5}
	\end{table}

	
	We transform the magnitude light-curves of 77,262 matched objects into normalized flux light-curves and extract 11 features (\emph{logP}, \emph{R$_{21}$}, \emph{R$_{31}$}, \emph{R$_{41}$}, $\sigma$, \emph{ALH}, \emph{M$_{s}$}, \emph{M$_{k}$}, \emph{MAD}, \emph{IQR}, \emph{A}) for each light curve. For the remaining 6 color-related features (\emph{G$_{BP}$} - \emph{G$_{RP}$}, \emph{J} - \emph{K$_{s}$}, \emph{J} - \emph{H}, \emph{W1} $-$ \emph{W2}, \emph{W$_{RP}$}, and \emph{W$_{JK}$}), we keep them unchanged. Excluding the aperiodic and feature-missing objects, a total of 60,082 variables have 17 features in both bands. 
	
	We use this classifier to classify the features of the 60,082 common objects in two bands. The classification results of 7 superclasses are shown in Table~\ref{tab:table6} with \texttt{CICD} = 0.897. For Eclipsing Binaries, $\delta$ Scuti, RR Lyrae variable stars, the classification numbers in the V and r bands are almost the same. The F$_{1}$* value of Cepheids is lower than 0.650, and the F$_{1}$* value of Mira is 0.701. For other types of variable stars, the values are above 0.829, and for Eclipsing Binaries, the value even reaches 0.942.
	
	The classification performance (i.e. F$_{1}$) of this classifier for each superclass of variable stars is above 0.822. The data in Table~\ref{tab:table6} show that Cepheids have significant classification differences, Mira, $\delta$ Scuti, ROT and RR Lyrae variables have moderate classification differences, Eclipsing Binaries and SR variables have minor classification differences.
	
	\begin{table}
		\centering
		\caption{Comparison of superclass classification results based on normalized flux light-curves.}
			\begin{tabular}{lccccc|c} 
				\hline
				Superclass & V-band number & r-band number & precision1* & precision2* & F$_{1}$* & CICD \\
				\hline
				Cepheids & 698 & 1261 & 0.759 & 0.420 & 0.541 & \multirow{7}{*}{0.897}\\
				$\delta$ Scuti & 2272 & 2202 & 0.816 & 0.842 & 0.829\\
				Eclipsing Binaries & 34558 & 35457 & 0.955 & 0.930 &0.942\\
				Mira & 1593 & 2527 & 0.907 & 0.572 & 0.701\\
				Rotational variables & 3275 & 2737 & 0.791 & 0.946 & 0.862\\
				RR Lyrae & 10098 & 9349 & 0.809  & 0.873 & 0.840\\
				Semiregular variables & 7588 & 6549 & 0.833 & 0.966 & 0.895\\
				\hline
			\end{tabular}
		\label{tab:table6}
	\end{table}
	
	The classification results for 17 subclasses are shown in Table~\ref{tab:table7} with \texttt{CICD} = 0.827. For CWA, ELL, RRD, and RVA variables, the classification numbers of this classifier in the V- and r-band are very small. Although the classification performance (i.e. F$_{1}$) for CWA, ELL and RVA variables is above 0.721, the F$_{1}$* values of these variables have less reference values.
	
	The F$_{1}$* values of CWB and RRC variables are 0.323 and 0.728, which appear to have significant and moderate classification differences, respectively. But the classification performance (i.e. F$_{1}$) for CWB and RRC variables is only 0.651 and 0.668, these results are for reference only.
	
	The classification performance (i.e. F$_{1}$) of the other 11 subclasses of variable stars is greater than 0.767, and can exceed 0.868 as a whole. The F$_{1}$* value of DCEPS variable stars is only 0.468, indicating that the DCEPS variables have significant classification differences. The F$_{1}$* values of DCEP, DSCT, EA, EB, HADS, Mira, and ROT variable stars range from 0.678 to 0.862, indicating that these variable stars have moderate classification differences. The precision1*, precision2*, and F$_{1}$* of EW, RRAB, and SR variables are all very high, and these results tend to indicate that EW, RRAB, and SR variables have minor classification differences.
	
	\begin{table}
		\centering
		\caption{Comparison of subclass classification results based on normalized flux light-curves.}
			\begin{tabular}{ccccccc|c} 
				\hline
				Superclass & Subclass & V-band number & r-band number & precision1* & precision2* & F$_{1}$* & CICD \\
				\hline
				\multirow{5}{*}{Cepheids} & CWA & 96 & 222 & 0.656 & 0.284 & 0.396 & \multirow{17}{*}{0.827}\\
				& CWB & 178 & 386 & 0.511 & 0.236	& 0.323\\
				& DCEP & 224 & 271 & 0.759 & 0.6279 & 0.687\\
				& DCEPS & 159 & 345 & 0.742 & 0.342 & 0.468\\
				& RVA & 41 & 37 & 0.585 & 0.649 & 0.615\\
				\cline{2-7}
				\multirow{2}{*}{$\delta$ Scuti} & DSCT & 692 & 926 & 0.825 & 0.617 & 0.706\\
				& HADS & 1580 & 1276 & 0.651 & 0.806 & 0.721\\
				\cline{2-7}
				\multirow{4}{*}{Eclipsing Binaries} & EA & 3811 & 4544 & 0.913 & 0.765 & 0.833\\
				& EB & 7164 & 7489 & 0.693 & 0.663 & 0.678\\
				& ELL & 57& 167 & 0.281 & 0.096 & 0.143\\
				& EW & 23526 & 23257 & 0.890 & 0.900 & 0.895\\
				\cline{2-7}
				\multirow{1}{*}{Mira} & M & 1593 & 2527 & 0.907 & 0.572	& 0.701\\
				\cline{2-7}
				\multirow{1}{*}{Rotational variables} & ROT & 3275 & 2737 & 0.791 & 0.946 & 0.862\\
				\cline{2-7}
				\multirow{3}{*}{RR Lyrae} & RRAB & 4499 & 4293 & 0.912 & 0.956 & 0.933\\
				& RRC & 5351 & 4899 & 0.698 & 0.762  & 0.728\\
				& RRD & 248 & 157 & 0.093 & 0.146 & 0.114\\
				\cline{2-7}
				\multirow{1}{*}{Rotational variables} & SR & 7588 & 6549 & 0.833 & 0.966 & 0.895\\
				\hline
			\end{tabular}
		\label{tab:table7}
	\end{table}
	
	In short, except for Eclipsing Binaries, EW, RRAB, and SR variables, other types of variable stars have notable classification differences. This random forest classifier is not good at predicting Cepheids and DCEPS varibales in different bands. For DCEP, DSCT, EB, HADS, and Mira variables in different bands, the classification results of this classifier need further verification. The classification results for $\delta$ Scuti, EA, Eclipsing Binaries, EW, ROT, RRAB, RR Lyrae, and SR variables are accurate with high probability.

	\subsection{The Classification Differences of Variables with Common Periods}
	\label{sec:packages3.2}
	
	By converting the magnitude light-curves into normalized flux light-curves, the consistency of classification type between two bands is indeed greatly improved, but the classification differences still exist. The period differences between V-band and r-band may be a factor resulting in the classification differences. To test the hypothesis, we try to classify light curves of different bands by using common periods. We first transform the magnitude light-curves of 77,262 matched objects into normalized flux light-curves. The time uses the HJD in ASAS-SN V-band. The time from ZTF DR2 data released by \cite{Chen2020} uses the HJD-2400000.5. Then we add 2400000.5 to the time of variables from ZTF DR2. According to the time, two normalized flux light-curves are combined to form the new flux light-curve.
	
	We calculate a common period for the new light curve and apply the period for both bands. For 75,924 objects with common periods, the new light curves are folded according to the common periods, and three common shape parameters (\emph{R$_{41}$}, \emph{R$_{31}$}, \emph{R$_{21}$}) are obtained by sixth-order Fourier series fitting. In addition to the above 4 features, other 7 features ($\sigma$, \emph{ALH}, \emph{M$_{s}$}, \emph{M$_{k}$}, \emph{MAD}, \emph{IQR}, \emph{A}) are obtained using single-band normalized flux light-curves, respectively. The remaining 6 color-related features (\emph{G$_{BP}$} - \emph{G$_{RP}$}, \emph{J} - \emph{K$_{s}$}, \emph{J} - \emph{H}, \emph{W1} $-$ \emph{W2}, \emph{W$_{RP}$}, and \emph{W$_{JK}$}) keep unchanged. Finally, there are 61,123 objects with all 17 features in both bands.				
	
	We use the random forest classifier to classify the features of 61,123 objects with common periods in two bands. The classification results for 7 superclasses are shown in Table~\ref{tab:table8} with \texttt{CICD} = 0.904. The classification results for 17 subclasses are shown in Table~\ref{tab:table9} with \texttt{CICD} = 0.850.
	
	\begin{table}
		\centering
		\caption{Comparison of superclass classification results based on normalized flux light-curves and with common periods.}
			\begin{tabular}{lccccc|c} 
				\hline
				Superclass & V-band number & r-band number & precision1* & precision2* & F$_{1}$* & CICD \\
				\hline
				Cepheids & 737 & 1201 & 0.786 & 0.482 & 0.598 & \multirow{7}{*}{0.904}\\
				$\delta$ Scuti & 2211 & 2154 & 0.850 & 0.873 & 0.861\\
				Eclipsing Binaries & 34696 & 35540 & 0.957 & 0.934 & 0.945\\
				Mira & 1725 & 2660 & 0.905 & 0.587 & 0.712\\
				Rotational variables & 3306 & 2800 & 0.800 & 0.945 & 0.866\\
				RR Lyrae & 9962 & 9328 &  0.823 & 0.879  & 0.850\\
				Semiregular variables & 8486  & 7440 & 0.853 & 0.973 & 0.909\\
				\hline
			\end{tabular}
		\label{tab:table8}
	\end{table}

	\begin{table}
		\centering
		\caption{Comparison of subclass classification results based on normalized flux light-curves and with common periods.}
			\begin{tabular}{ccccccc|c} 
				\hline
				Superclass & Subclass & V-band number & r-band number & precision1* & precision2* & F$_{1}$* & CICD \\
				\hline
				\multirow{5}{*}{Cepheids} & CWA & 106 & 213 & 0.679 & 0.338 & 0.451 & \multirow{17}{*}{0.850}\\
				& CWB & 186 & 363 & 0.554  & 0.284 & 0.375\\
				& DCEP & 238 & 261 & 0.782 & 0.713 & 0.745\\
				& DCEPS & 167 & 325 & 0.784 & 0.403 & 0.533\\
				& RVA & 40 & 39 & 0.700 & 0.718 & 0.709\\
				\cline{2-7}
				\multirow{2}{*}{$\delta$ Scuti} & DSCT & 684 & 922 & 0.832 & 0.617 & 0.709\\
				& HADS & 1527 & 1232	& 0.688  & 0.852 & 0.761 \\
				\cline{2-7}
				\multirow{4}{*}{Eclipsing Binaries} & EA & 3904 & 4344 & 0.975 & 0.877 & 0.923\\
				& EB & 6989 & 7698 & 0.758 & 0.688 & 0.721\\
				& ELL & 77 & 180 & 0.597 & 0.256 & 0.358\\
				& EW & 23726 & 23318 & 0.893 & 0.909 & 0.901\\
				\cline{2-7}
				\multirow{1}{*}{Mira} & M & 1725 & 2660	& 0.905 & 0.587 & 0.712\\
				\cline{2-7}
				\multirow{1}{*}{Rotational variables} & ROT & 3306 & 2800 & 0.800 & 0.945 & 0.866\\
				\cline{2-7}
				\multirow{3}{*}{RR Lyrae} & RRAB & 4393 & 4216 & 0.944 & 0.984 & 0.964\\
				& RRC & 5352 & 4968 & 0.716 & 0.771 & 0.743\\
				& RRD & 217 & 144 & 0.088 & 0.132 & 0.105\\
				\cline{2-7}
				\multirow{1}{*}{Semiregular variables} & SR & 8486 & 7440 & 0.853 & 0.973 & 0.909\\
				\hline
			\end{tabular}
		\label{tab:table9}
	\end{table}

	From these results, we can see that when the normalized flux light-curves of two bands are combined to obtain common periods, except for Eclipsing Binaries, EA, EW, RRAB, and SR variables, other types of variable stars have notable classification differences. This classifier is not good at predicting Cepheids and DCEPS variables in different bands. For different bands of DCEP, DSCT, EB, HADS, and Mira variables, the classification results of this classifier need further verification. The classification results for $\delta$ Scuti, EA, Eclipsing Binaries, EW, ROT, RRAB, RR Lyrae, and SR variables are accurate with high probability.
	
\begin{figure*}
		\centering
		\includegraphics[width=\textwidth]{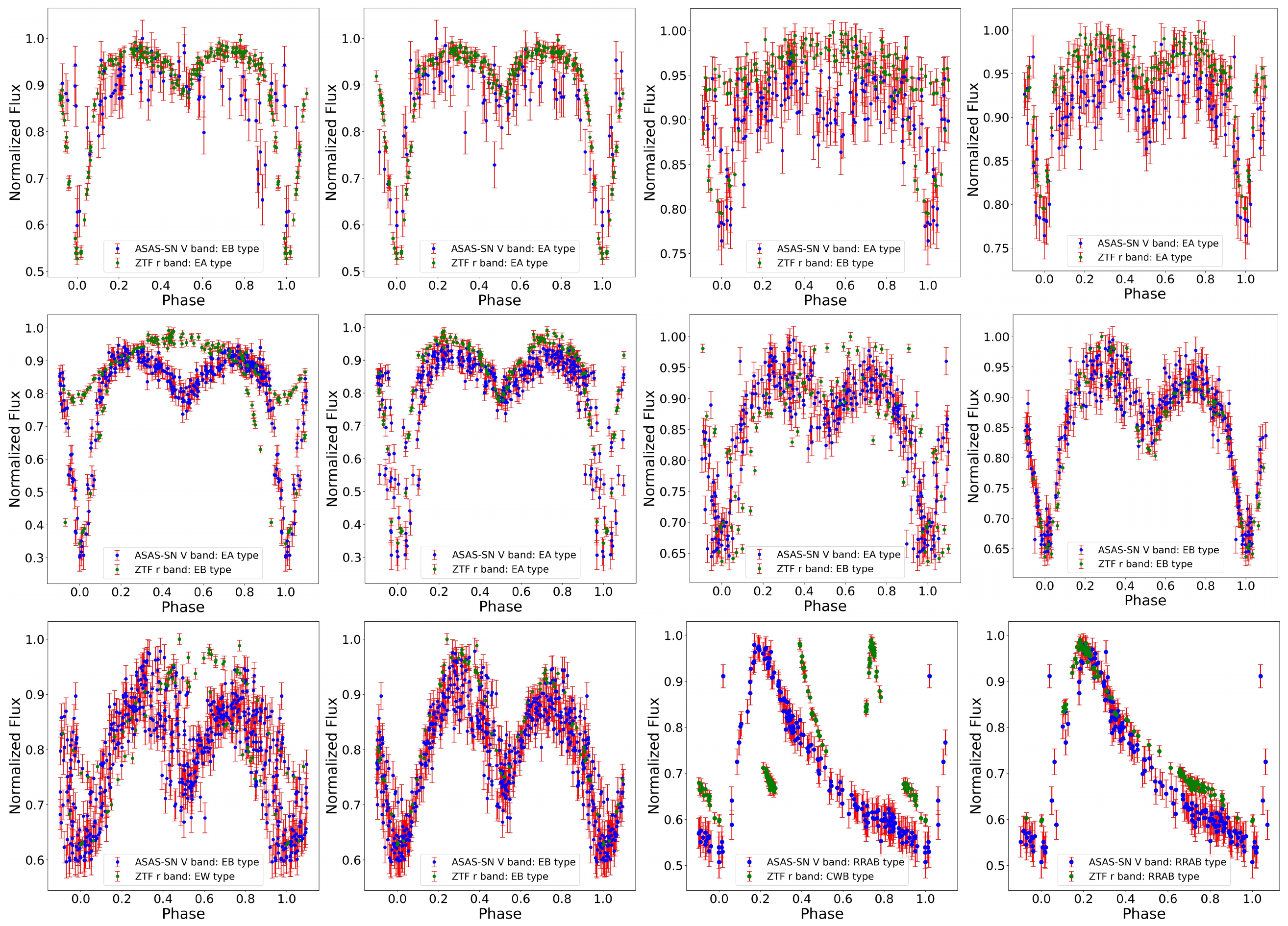}
		\caption{Classification of several variables in normalized flux light curves from the ASAS-SN V band and ZTF DR2 r band. When periods are derived from single-band light curves, the classification types are shown in the first and third columns. When periods are derived from combined multi-band light curves, the classification types are presented in the second and fourth columns.}
		\label{fig:figure1}
	\end{figure*}

	Compared with the data in Table~\ref{tab:table6}, the data in Table~\ref{tab:table8} have a slight improvement. Compared with the data in Table~\ref{tab:table7}, the values of the three parameters of ELL and RVA variables in Table~\ref{tab:table9} vary greatly, but the F$_{1}$* values of the two types of variable stars both have less reference values. For EA variable stars, the F$_{1}$* has increased by $\sim$ 10\%. These comparisons show that with color-related features, for most types of variable stars, the periods calculated by the V- and r-band light curves are almost the same. However, for EA variables, there are some differences in the periods calculated by the light curves of different bands. Fig.~\ref{fig:figure1} shows the classification types of several variables in the normalized flux light curves of ASAS-SN V band and ZTF r band. Combining the light curves of multiple bands to obtain the common periods is conducive to improving the consistency of classification types.

	\subsection{The Classification Differences without Color-related Features}
	\label{sec:packages3.3}
	There are classification differences in the case of magnitude light-curves to classify variable stars in the V- and r-band. The classification differences can not be completely eliminated by converting magnitude light-curves into the normalized flux light-curves or combining the normalized flux light-curves to obtain the common periods. Considering that the color-related features may bring the classification differences and retaining only objects with color-related features significantly reduces sample size, we exclude the 6 color-related features (\emph{G$_{BP}$} - \emph{G$_{RP}$}, \emph{J} - \emph{K$_{s}$}, \emph{J} - \emph{H}, \emph{W1} $-$ \emph{W2}, \emph{W$_{RP}$}, and \emph{W$_{JK}$}), and only retain the 11 features (\emph{logP}, \emph{R$_{41}$}, \emph{R$_{31}$}, \emph{R$_{21}$}, $\sigma$, \emph{ALH}, \emph{M$_{s}$}, \emph{M$_{k}$}, \emph{MAD}, \emph{IQR}, \emph{A}) extracted from the normalized flux light-curves of 235,491 objects.We use 80\% of the features for training and 20\% of the features for testing. Finally, we use all the features to retrain the random forest classifier.
	
	Table~\ref{tab:table10} shows the precision, recall, and F$_{1}$ of this classifier in superclasses. Compared with Table~\ref{tab:table4}, the classification performance (i.e. F$_{1}$) on $\delta$ Scuti and ROT variables is reduced significantly and decreased slightly for Cepheids, and SR variables, indicating that adding color-related information is beneficial for distinguishing these types of variables. Table~\ref{tab:table11} shows the precision, recall, and F$_{1}$ on subclasses. Due to the small numbers of CWA, CWB, RRD, RVA variables during testing the classification performance, we do not compare with Table~\ref{tab:table5} for these types of variables. The classification performance (i.e. F$_{1}$) for DSCT variables decreases obviously and decreases slightly for DCEP, DCEPS, ELL, HADS, and RRC variables. It shows that adding color-related information is helpful to distinguish these subclasses of variable stars. We note that the recall of RRD variables increase by 14\%, but the precision decrease by $\sim$ 12\%, indicating that the removal of color-related information is beneficial for true RRD variables to accurately classify, but it also leads to more other variables being misclassified as RRD variables. This classifier continues to perform well in the classification of Eclipsing Binaries, EA, EB, EW, Mira, RR Lyrae, and RRAB variable stars, indicating that color-related information is not unique characteristic information of these types of variable stars and can not distinguish them from other types of variable stars.

	\begin{table}
		\centering
		\caption{The classification quality of superclasses without color-related features.}
			\begin{tabular}{lccc} 
				\hline
				Superclass & precision & recall & F$_{1}$\\
				\hline
				Cepheids & 0.812 & 0.951 & 0.876\\
				$\delta$ Scuti & 0.572 & 0.943 &0.712\\
				Eclipsing Binaries & 0.976 & 0. 905 & 0.939\\
				Mira & 0.902 & 0.987 & 0.943\\
				Rotational variables & 0.597 & 0.837 & 0.697\\
				RR Lyrae & 0.826 & 0.944 & 0.881\\
				Semiregular variables & 0.989 & 0.855 & 0.917\\
				\hline
			\end{tabular}
		\label{tab:table10}
	\end{table}
	
	
	\begin{table}
		\caption{The classification quality of subclasses without color-related features.}
			\centering
			\begin{tabular}{ccccc} 
				\hline
				Superclass & Subclass & precision & recall & F$_{1}$\\
				\hline
				\multirow{5}{*}{Cepheids} & CWA & 0.502 & 0.860 & 0.634\\
				& CWB	& 0.389 & 0.831 & 0.530\\
				& DCEP & 0.908 & 0.812 & 0.857\\
				& DCEPS & 0.783 & 0.903 & 0.839\\
				& RVA	& 0.556 & 0.959  & 0.704\\
				\cline{2-5}
				\multirow{2}{*}{$\delta$ Scuti} & DSCT & 0.469 & 0.981 & 0.635\\
				& HADS & 0.663 & 0.893 & 0.761\\
				\cline{2-5}
				\multirow{4}{*}{Eclipsing Binaries} & EA & 0.962 & 0.919 & 0.940\\
				& EB & 0.729 & 0.785 & 0.756\\
				& ELL	& 0.851 & 0.909 & 0.879\\
				& EW	& 0.931 & 0.771 & 0.843\\
				\cline{2-5}
				\multirow{1}{*}{Mira} & M & 0.902 & 0.987 & 0.943\\
				\cline{2-5}
				\multirow{1}{*}{Rotational variables} & ROT	& 0.597 & 0.837 & 0.697\\
				\cline{2-5}
				\multirow{3}{*}{RR Lyrae} & RRAB & 0.982 & 0.952 & 0.967\\
				& RRC	&  0.520 & 0.787 & 0.626\\
				& RRD	&0.211 &  0.761 & 0.330\\
				\cline{2-5}
				\multirow{1}{*}{Semiregular variables} & SR & 0.989 & 0.855 & 0.917 \\
				\hline
			\end{tabular}
		\label{tab:table11}
	\end{table}
	

	Compared to the 61,123 common sources in Section~\ref{sec:packages3.2}, as we do not consider color-related information, more variable stars are retained. A total of 75,924 objects have 11 features based on normalized flux light-curves in both bands.
	
	Similarly, we use this classifier to classify the features of 75,924 common sources in two bands. The classification results of the 7 superclasses are shown in Table~\ref{tab:table12} with \texttt{CICD} = 0.850. The classification results of 17 subclasses are shown in Table~\ref{tab:table13} with \texttt{CICD} = 0.788.
	
	\begin{table}
		\centering
		\caption{Comparison of superclass classification results without color-related features.}
			\begin{tabular}{lccccc|c} 
				\hline
				Superclass & V-band number & r-band number & precision1* & precision2* & F$_{1}$*& CICD \\
				\hline
				Cepheids & 1832 & 3048 & 0.743 & 0.447 & 0.558 & \multirow{7}{*}{0.850}\\
				$\delta$ Scuti & 3897 & 4429 & 0.723 & 0.636 & 0.677\\
				Eclipsing Binaries & 36193 & 36751 & 0.904 & 0.890 & 0.897\\
				Mira & 1973 & 3153 & 0.900 & 0.563 & 0.693\\
				Rotational variables & 6571  & 5202 & 0.690 & 0.872 & 0.771\\
				RR Lyrae & 16102 & 15306 & 0.851 & 0.896 & 0.873\\
				Semiregular variables & 9356 & 8035 & 0.812 & 0.945 & 0.873\\
				\hline
			\end{tabular}
		\label{tab:table12}
	\end{table}

	\begin{table}
		\centering
		\caption{Comparison of subclass classification results without color-related features.}
			\begin{tabular}{ccccccc|c} 
				\hline
				Superclass & Subclass & V-band number & r-band number & precision1* & precision2* & F$_{1}$* & CICD \\
				\hline
				\multirow{5}{*}{Cepheids} & CWA & 173 & 520 & 0.705 & 0.235 & 0.352 & \multirow{17}{*}{0.788}\\
				& CWB & 807 & 928 & 0.538 & 0.468 & 0.500\\
				& DCEP & 415 & 615 & 0.646 & 0.436 & 0.520\\
				& DCEPS & 369 & 916 & 0.518 & 0.209 & 0.297\\
				& RVA & 68 & 69 & 0.279 & 0.275 & 0.277\\
				\cline{2-7}
				\multirow{2}{*}{$\delta$ Scuti} & DSCT & 1316 & 1903 & 0.787 & 0.544 & 0.644\\
				& HADS & 2581 & 2526 & 0.508 & 0.519 & 0.513\\
				\cline{2-7}
				\multirow{4}{*}{Eclipsing Binaries} & EA & 4403 & 4851 & 0.969 & 0.880 & 0.922\\
				& EB & 7257 & 8125 & 0.743 & 0.664 & 0.701\\
				& ELL & 481 & 955 & 0.690 & 0.348 & 0.462\\
				& EW & 24052 & 22820 & 0.814 & 0.857 & 0.835\\
				\cline{2-7}
				\multirow{1}{*}{Mira} & M & 1973 & 3153 & 0.900 & 0.563 & 0.693\\
				\cline{2-7}
				\multirow{1}{*}{Rotational variables} & ROT & 6571 & 5202 & 0.690 & 0.872 & 0.771\\
				\cline{2-7}
				\multirow{3}{*}{RR Lyrae} & RRAB & 9789 & 9658 & 0.960 & 0.973 & 0.966\\
				& RRC & 5612 & 5033 & 0.622 & 0.694 & 0.656\\
				& RRD & 701 & 615 & 0.143 & 0.163 & 0.152\\
				\cline{2-7}
				\multirow{1}{*}{Semiregular variables} & SR & 9356 & 8035 & 0.812 & 0.945 & 0.873\\
				\hline
			\end{tabular}
		\label{tab:table13}
	\end{table}

	In the absence of color-related features, in the case of combining normalized flux light-curves to obtain common periods, except for Eclipsing Binaries, EA, and RRAB variables, other types of variable stars have notable classification differences. This random forest classifier is not good at predicting Cepheids, DCEP, DCEPS, ELL, and HADS variables in different bands. For $\delta$ Scuti, EB, Mira, and ROT variables in different bands, the classification results of this classifier need further verification. The classification results for EA, Eclipsing Binaries, EW, RRAB, RR Lyrae, and SR variables are accurate with high probability. 
	
\begin{figure*}
		\centering
		\includegraphics[width=\textwidth]{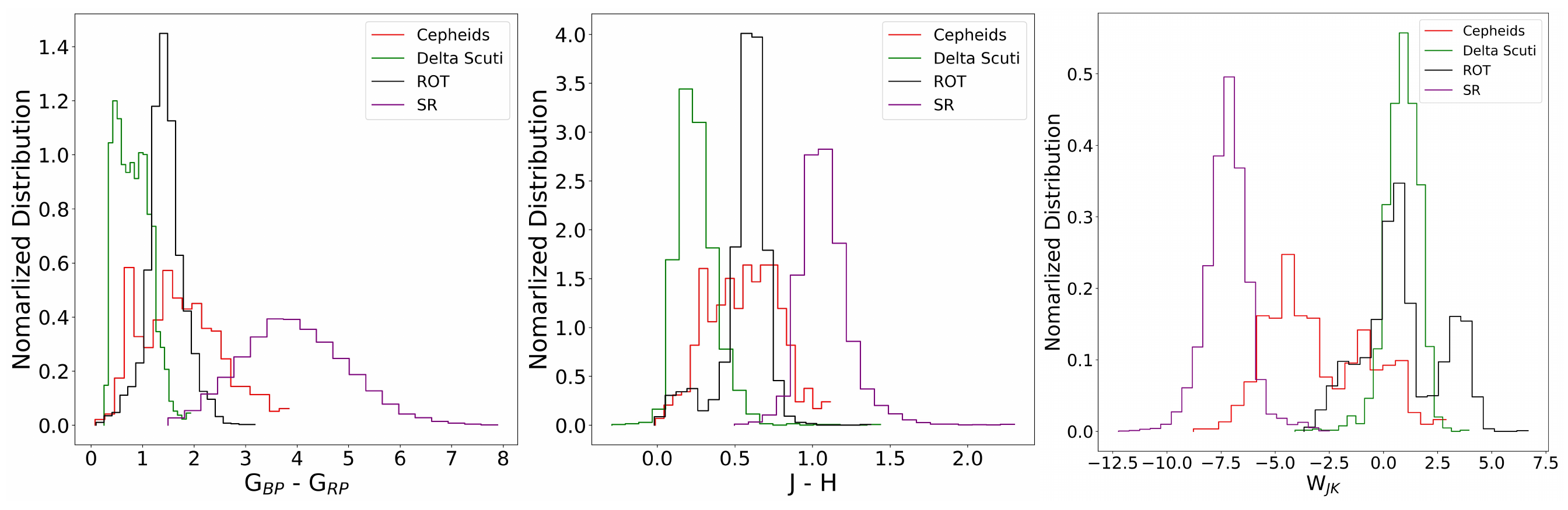}
		\caption{The normalized distributions of some color-related features for 12,020 objects which have the common periods and the same classification types. The red, green, black, and purple lines represent the Cepheids, $\delta$ Scuti, ROT, and SR variables, respectively.}
		\label{fig:figure2}
	\end{figure*}

	Compared with Table~\ref{tab:table8} and Table~\ref{tab:table9}, the values of \texttt{CICD} in Table~\ref{tab:table12} and Table~\ref{tab:table13} decrease by about 5\%. Without color-related features, under the condition of combining normalized flux light-curves to obtain the common periods, the random forest classifier's classification consistency (i.e. F$_{1}$*) in different bands for DCEP, DCEPS, $\delta$ Scuti, HADS, and ROT variables is obviously reduced by 12.5\%, 23.6\%, 18.4\%, 24.8\%, and 9.5\%, respectively. Color-related information is actually very important and helpful for distinguishing these types of variable stars in different bands. The random forest classifier's classification consistency (i.e. F$_{1}$*) in different bands for EA, EB, Mira, RRAB, RR Lyrae variables is almost unchanged, which indicates that color-related information for these variables can not help to improve the classification consistency under the condition of combining normalized flux light-curves to obtain the common periods,
	
	Fig.~\ref{fig:figure2} shows the normalized distributions of three color-related features for 12,020 objects which have the common periods and the same classification types under the conditions of normalized flux light-curves and considering color-related features. The 11 features extracted from the V- and r-band light cures can mostly distinguish the Mira variables (such as the \emph{A} and the \emph{period} features), Eclipsing Binaries and RR Lyraes (such as the \emph{ALH} and the \emph{$\sigma$} features) \citep{Xu2022}. For Cepheids, $\delta$ Scuti, ROT, SR variables, and their subclasses, with the help of one or two color-related features, they can be distinguished better. To be true, the color-related features are also helpful for other types of variables \citep{Xu2022}. As for why color-related features are important in distinguishing these types of variable stars in astrophysics, more researches needs to be done in the future.                
	
\section{Discussions}
\label{sec:packages4}
    From the above results, we note that only RRAB variables have minor classification differences in two bands under the condition of magnitude light-curves. If we use normalized flux light-curves, Eclipsing Binaries, EW, RRAB and SR variables have minor classification differences. When we use the normalized flux light-curves and the periods derived from the combined multi-band light-curves, Eclipsing Binaries, EA, EW, RRAB and SR variables have minor classification differences. Without the related-color features, Eclipsing Binaries, EA, RRAB variables have minor classification differences. It is possible to apply a same classifier to classify these types of variables across different bands. The differences between V and r bands can not be negligible for other types of variables, especially Cepheids. 
    
    Nevertheless, there are still many issues to be considered. As stated in Section~\ref{sec:packages1}, there are evidences suggesting that similar variability exists in different bands for many types of variables. One generally assumes that variability properties across different bands are correlated and the shapes of the light curves of different bands would remain consistent and magnitudes of different bands would vary synchronously. Such assumptions lead our attempt to develop a classifier that can be used in multiple bands.

    It is obvious that magnitudes of different bands generally take amplitudes within different ranges \citep{2010ApSS.326..219T,2022ApJS..260...46I}. To avoid the possible effects in classifiers caused by amplitude differences in magnitude light curves, we use normalized flux light curves to train the classifier. We find that the classification performance (i.e. F$_{1}$) based on magnitude light-curves (see Tables 7, 8 in \cite{Xu2022}) is generally better than that based on normalized flux light-curves (see Tables~\ref{tab:table4}, \ref{tab:table5}). The classification consistency based on normalized flux light-curves (i.e. F$_{1}$*, see Tables~\ref{tab:table14}, ~\ref{tab:table15} and Tables~\ref{tab:table6}, ~\ref{tab:table7}) is higher for all types. This suggests that the classifier trained by the single band magnitude light-curves is more beneficial for identifying and classifying the light curves in the same band. The normalized flux light-curves are useful to reduce classification differences and the classifier trained by normalized flux light-curves is more efficient to recognize and classify the light curves from different bands. Even though using normalized flux light-curves improves the classification consistency of Cepheids and their subclasses, it is not sufficiently effective. This may indicate the intrinsic shape differences of light curves across different bands. \cite{Kessler2019} revealed the light curve models of the transient sources and variable stars in ugrizy bands. Their data demonstrated that the flux range, amplitude and shape of light curves in different bands for some objects are different.

    Based on the comparisons between Tables~\ref{tab:table6}, ~\ref{tab:table7} and Tables~\ref{tab:table8}, ~\ref{tab:table9}, we suggest that for most types of variable stars, the periods calculated by V- and r- band light curves are almost the same. But for EA variables, the periods calculated by single band light curves are sometimes different. Therefore, combining multi-band light curves to obtain the common periods is conducive to improving the consistency of classification types, as shown in Fig.~\ref{fig:figure1}. 
    
    The color values are assumed to stay constant for all objects since we obtain single color measurements for each object from Gaia, 2MASS, and ALLWISE. These colors are observed at random phases within the periods of the variables. Stellar activity, such as due to starspots or pulsations \citep{2009AARv..17..251S,2010aste.book.....A}, can cause the color of a star to vary over time. Therefore, the assumption of constant colors is not justified. In our work, according to the comparison of classification consistency with and without color-related features (i.e. F$_{1}$*, see Tables~\ref{tab:table8}, ~\ref{tab:table9} and Tables~\ref{tab:table12}, ~\ref{tab:table13}), color-related information is crucial for distinguishing DCEP, DCEPS, $\delta$ Scuti, HADS, and ROT variables. 
    For other types of variable stars, adding color information does not aid in identification and classification. As shown in figure~\ref{fig:figure2}, adding color information still helps when the time-varying colors have distinguishable distributions. Moreover, it is important to note that many samples currently lack color information, so the decision to use color information as a feature for training classifiers should be made cautiously. 
    
    Additionally, a classifier trained on V-band photometric data may suffer from band bias when applied to V- and R-band samples, leading to more consistent classification results only within the same band as the training data. To avoid this bias, it is best to test in different bands from that of the training samples and avoid overlap between the training and testing samples. Adding a small number of OGLE samples to train the classifier may also impact its performance. Completely mitigating the effects of sample selection and imbalance remains a challenge for the future. More theoretical studies and observational data will be needed to validate the variability characteristics across different bands in the future.

\section{Conclusions}
\label{sec:packages5}
    We construct the random forest classifier based on variable stars from ASAS-SN V-band and OGLE I-band, and use the progressive  random forest classifier to classify ASAS-SN V- and ZTF r-band light curves, to explore whether there are differences in multi-band classification of variable stars. 
    
    By comparing the classification results using the magnitude light-curves and the normalized flux light-curves, the periods derived from single band normalized flux light-curves and the periods derived from the combined multi-band normalized flux light curves, with and without the color-related features, we conclude that using normalized flux light-curves, obtaining common periods, and incorporating color-related information are more conducive to recognizing and classifying multi-band variable stars. The classifier can be used in both bands for RRAB variables under the condition of magnitude light curves. Using normalized flux light curves, the classifier can be applied to Eclipsing Binaries, EW, RRAB, and SR variables. In the case of periods derived from combined multi-band light curves, the classifier works for Eclipsing Binaries, EA, EW, RRAB, and SR variables. Without color-related features, the classifier is effective for Eclipsing Binaries, EA, and RRAB variables. 
    
    In the current situation, one should be careful when applying a classifier trained with a band A data on a band B data. Although the classification differences for some types of variable stars cannot be excluded using these methods, through more extensive and in-depth research in the future, the development of a classifier that can be used across multiple bands is something to look forward to.

	\begin{acknowledgements}
		This work is supported by the Strategic Priority Research Program of Chinese Academy of Sciences (No. XDB 41000000), Anhui Provincial Natural Science Foundation (2308085QA35), China Postdoctoral Science Foundation (2021M703099), and the Fundamental Research Funds for the Central Universities. The authors gratefully acknowledge the support of Cyrus Chun Ying Tang Foundations.
		
		This work has made use of data from the ASAS-SN Variable Stars Database (\url{https://asas-sn.osu.edu/variables}). ASAS-SN is funded in part by the Gordon and Betty Moore Foundation through grants GBMF5490 and GBMF10501 to the Ohio State University, and also funded in part by the Alfred P. Sloan Foundation grant G-2021-14192. TJ, KZS and CSK are supported by NSF grants AST-1814440 and AST-1908570. B.J.S. is supported by NASA grant 80NSSC19K1717 and NSF grants AST-1920392 and AST-1911074. Support for T.W.-S.H. was provided by NASA through the NASA Hubble Fellowship grant HST-HF2-51458.001-A awarded by the Space Telescope Science Institute (STScI), which is operated by the Association of Universities for Research in Astronomy, Inc., for NASA, under contract NAS5-26555. S.D. acknowledges Project 12133005 supported by National Natural Science Foundation of China (NSFC). 
		
		Development of ASAS-SN has been supported by NSF grant AST-0908816, the Mt. Cuba Astronomical Foundation, the Center for Cosmology and AstroParticle Physics at the Ohio State University, the Chinese Academy of Sciences South America Center for Astronomy (CAS-SACA), the Villum Foundation, and George Skestos. TAT is supported in part by Scialog Scholar grant 24216 from the Research Corporation. Support for JLP is provided in part by FONDECYT through the grant 1151445 and by the Ministry of Economy, Development, and Tourism’s Millennium Science Initiative through grant IC120009, awarded to The Millennium Institute of Astrophysics, MAS.
		
		This work has made use of data from the OGLE Collection of Variable Stars (\url{https://ogledb.astrouw.edu.pl}). The OGLE project has received funding from the European Research Council under the European Community’s Seventh Framework Programme (FP7/2007-2013) / ERC grant agreement no. 246678 to AU. RP is supported by the Foundation for Polish Science through the Start Program.
		
		This publication is based on observations obtained with the Samuel Oschin 48-inch Telescope at the Palomar Observatory as part of the Zwicky Transient Facility project. ZTF is supported by the National Science Foundation under grant AST-1440341 and a collaboration including Caltech, IPAC, the Weizmann Institute for Science, the Oskar Klein Center at Stockholm University, the University of Maryland, the University of Washington, Deutsches Elektronen-Synchrotron and Humboldt University, Los Alamos National Laboratories, the TANGO Consortium of Taiwan, the University of Wisconsin at Milwaukee, and Lawrence Berkeley National Laboratories. Operations are conducted by COO, IPAC, and UW.
		
		This work has made use of data from the European Space Agency (ESA) mission Gaia (\url{https://www.cosmos.esa.int/gaia}), processed by the Gaia Data Processing and Analysis Consortium (DPAC, \url{https://www.cosmos.esa.int/web/gaia/dpac/consortium}). Funding for the DPAC has been provided by national institutions, in particular the institutions participating in the Gaia Multilateral Agreement.
		
		This publication makes use of data products from the Widefield Infrared Survey Explorer, which is a joint project of the University of California, Los Angeles, and the Jet Propulsion Laboratory/California Institute of Technology, funded by the National Aeronautics and Space Administration. 
		
		This publication makes use of data products from the Two Micron All Sky Survey, which is a joint project of the University of Massachusetts and the Infrared Processing and Analysis Center/California Institute of Technology, funded by the National Aeronautics and Space Administration and the National Science Foundation.
		
		~\\
		\emph{Software}: Scikit-learn \citep{Pedregosa2012}, Astropy \citep{AstropyCollaboration2013}, Astrobase \citep[v0.5.3,][]{Bhatti2021}, Gatspy \citep[v0.3,][]{Vanderplas2016}.
	\end{acknowledgements}

	\appendix                  
	
	\section{The Classification Differences Based on Magnitude Light-Curves}
	\label{sec:packages6}
	We use the existing random forest classifier trained by \cite{Xu2022} to investigate the classification differences in multi-band magnitude light curves. We use the period determination method proposed by \cite{Xu2022} to calculate periods for 77,262 sources in V and r bands and extract 17 features for each magnitude light curve of the two bands. Excluding non-periodic and feature-missing sources, a total of 53,851 objects have 17 features in both bands. We use the classifier to classify the features of 53,851 common objects in two bands. The classification results of this classifier for 7 superclasses are shown in Table~\ref{tab:table14} with \texttt{CICD} = 0.777. For Eclipsing Binaries and $\delta$ Scuti variables, the classification numbers in the V-band and r-band are roughly the same. Although the classification performance (i.e. F$_{1}$) for each superclass is above 0.931, the values of F$_{1}$* for all types of variable stars are between 0.359 and 0.845. It suggests that classification differences between two bands exist. Cepheids have significant classification differences, while Eclipsing Binaries have high classification consistency.

	\begin{table}
		\centering
		\caption{Comparison of superclass classification results based on magnitude light-curves.}
			\begin{tabular}{lccccc|c} 
				\hline
				Superclass & V-band number & r-band number & precision1* & precision2* &  F$_{1}$* & CICD \\
				\hline
				Cepheids & 728 & 1521 & 0.555 & 0.266 & 0.359 & \multirow{7}{*}{0.777}\\
				Delta Scuti & 5987 & 5840 & 0.652 & 0.668 & 0.660\\
				Eclipsing Binaries & 30172	& 31099 & 0.858 & 0.832 & 0.845\\
				Mira & 1561	& 2378 & 0.881 & 0.578 & 0.698\\
				Rotational variables & 1897 & 1504 &0.690 & 0.870 & 0.769\\
				RR Lyrae & 9622	& 8494 & 0.688 & 0.779 & 0.731\\
				Semiregular variables & 3884 & 3015 & 0.608 & 0.783 & 0.685\\
				\hline
			\end{tabular}
		\label{tab:table14}
	\end{table}

	The classification results of this classifier for 17 subclasses are shown in Table~\ref{tab:table15} with \texttt{CICD} = 0.716. For CWA, CWB, RRD, RVA variable stars, the classification numbers in the V-band and r-band are too small, their results have less reference values.
	
	The classification performance (i.e. F$_{1}$) of other 13 subclasses of variable stars is above 0.799. For DCEP, DCEPS, DSCT, ELL, HADS and RRC variable stars, their F$_{1}$* values are all below 0.650, indicating that these variable stars have significant classification differences. The F$_{1}$* values of the remaining 7 subclasses of variable stars are above 0.700 as a whole, and the value for RRAB variables even reaches 0.912, indicating that EA, EB, EW, Mira, ROT, and SR variables also have the classification differences, but the differences are moderate compared with the 6 subclasses of variable stars mentioned above. The precision1*, precision2*, and F$_{1}$* are all high for RRABs, which tend to indicate that RRAB variables have minor classification differences.

	\begin{table}
		\centering
		\caption{Comparison of subclass classification results based on magnitude light-curves.}
			\begin{tabular}{ccccccc|c} 
				\hline
				Superclass & Subclass & V-band number & r-band number & precision1* & precision2* & F$_{1}$* & CICD \\
				\hline
				\multirow{5}{*}{Cepheids} & CWA & 96 & 204 & 0.583 & 0.275 & 0.373 & \multirow{17}{*}{0.716}\\
				& CWB & 128 & 320 & 0.578 & 0.231	& 0.330\\
				& DCEP & 291 & 633 & 0.316 & 0.145 & 0.199\\
				& DCEPS & 160 & 326 & 0.631 & 0.310 & 0.416\\
				& RVA & 53 & 38 & 0.415 & 0.579 & 0.484\\
				\cline{2-7}
				\multirow{2}{*}{Delta Scuti} & DSCT & 791 & 1018 & 0.632 & 0.491 & 0.553\\
				& HADS & 5196 & 4822 & 0.582 & 0.627 & 0.604\\
				\cline{2-7}
				\multirow{4}{*}{Eclipsing Binaries} & EA & 3543 & 4532 & 0.909 & 0.711 & 0.798\\
				& EB & 6892 & 6595 & 0.703 & 0.735 & 0.719\\
				& ELL & 775 & 604 & 0.543 & 0.697 & 0.611\\
				& EW & 18962 & 19368 & 0.774 & 0.758 & 0.766\\
				\cline{2-7}
				\multirow{1}{*}{Mira} & M & 1561 & 2378 & 0.881 & 0.578	& 0.698\\
				\cline{2-7}
				\multirow{1}{*}{Rotational variables} & ROT & 1897 & 1504 & 0.690 & 0.870 & 0.769\\
				\cline{2-7}
				\multirow{3}{*}{RR Lyrae} & RRAB & 4293 & 3875 & 0.867 & 0.961 & 0.912\\
				& RRC & 5256 & 4529 & 0.524 & 0.608 & 0.562\\
				& RRD & 73 & 90 & 0.096 & 0.078 & 0.086\\
				\cline{2-7}
				\multirow{1}{*}{Semiregular variables} & SR & 3884 & 3015 & 0.608 & 0.783 & 0.685\\
				\hline
			\end{tabular}
		\label{tab:table15}
	\end{table}

\begin{figure*}
		\centering
		\includegraphics[width=\textwidth]{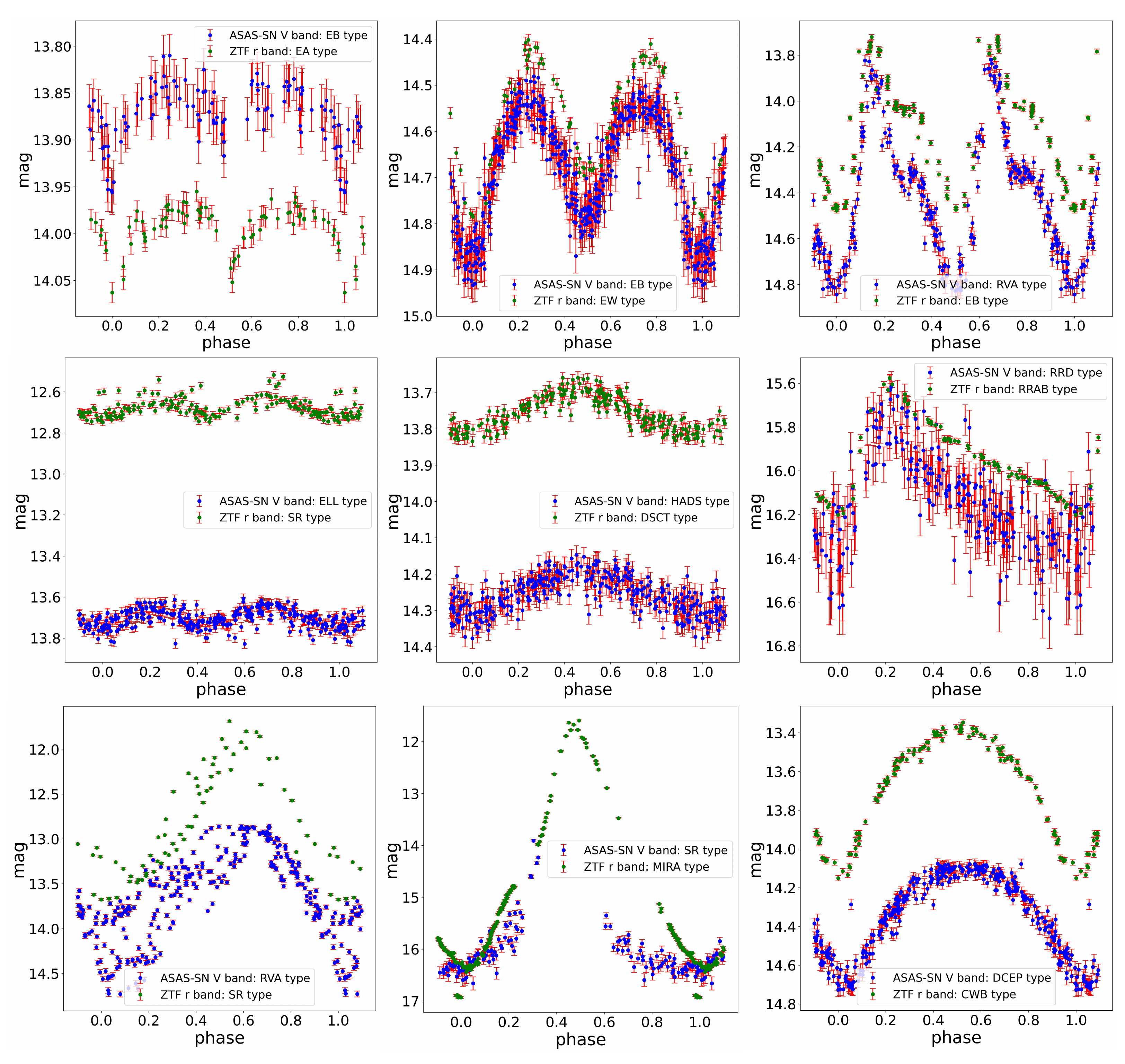}
		\caption{The different classification of several variables in the magnitude light curves of ASAS-SN V band and ZTF r band.}
		\label{fig:figure3}
	\end{figure*}

	We show the classification types of several variables in ASAS-SN V band and ZTF r band in Fig.~\ref{fig:figure3}.  Due to the differences in magnitude and amplitude, misclassification is more likely to occur between EA, EB, EW variables, and between HADS and DSCT.
	
	In summary, except for RRAB variables, other types of variable stars have notable classification differences in two bands. This random forest classifier is not good at predicting types for Cepheids, DCEP, DCEPS, DSCT, ELL, HADS, RRC variables in different bands. The classification results in different bands for $\delta$ Scuti, EB, EW, Mira, ROT, RR Lyrae, SR variable stars need further verification. For Eclipsing Binaries, EA, and RRAB variables, the classification results are accurate with high probability.

\end{document}